%% file: iceprod.tex
\documentclass[longtitle,preprint,3p,times,twocolumn]{elsarticle}
\usepackage{lineno}
\usepackage{amssymb}
\usepackage{graphicx}
\usepackage{url}
\usepackage{color}
\usepackage{perpage} 
\usepackage{caption} 

\MakePerPage{footnote} 

\journal{Journal of Parallel and Distributed Computing}
\begin{document}

\begin{frontmatter}
\title{The IceProd Framework: \\
Distributed Data Processing for the IceCube Neutrino Observatory}


\hyphenation{distributed identify IceProd detector order resources daemons}
\include{authors-elsv}

\begin{abstract}
IceCube is a one-gigaton 
instrument located at the geographic South Pole, designed to detect cosmic neutrinos, identify the particle nature of dark matter, and 
study high-energy neutrinos themselves.
Simulation of the IceCube detector and processing of data require a
significant amount of computational resources. This paper presents the first detailed description of IceProd, a lightweight distributed management system 
designed to meet these requirements. It is driven by a central database in order to
manage mass production of simulations and analysis of data produced by the
IceCube detector. IceProd runs as a separate layer on top of other middleware and can take advantage of a variety of
computing resources, including grids and batch systems such as CREAM, HTCondor, and PBS.
This is accomplished by a set of dedicated daemons that process job submission in a 
coordinated fashion through the use of middleware plugins that serve to abstract the 
details of job submission and job management from the framework. 
\end{abstract}

\begin{keyword}
Data Management, Grid Computing, Monitoring, Distributed Computing
\end{keyword}

\end{frontmatter}


%
%
\section{Introduction}
\label{ch:intro}

Large experimental collaborations often need to produce  extensive volumes
of computationally intensive Monte Carlo simulations and process vast amounts of data. 
These tasks are usually farmed out to large computing clusters or
grids. For such large datasets, it is important to be able to
document details associated with each task, such as software versions and parameters like the 
pseudo-random number generator seeds used for each dataset. Individual members of
such collaborations might have access to modest computational
resources that need to be coordinated for production. Such computational 
resources could also potentially be pooled in order to provide a single, more powerful, and more
productive system that can be used by the entire collaboration. 
This article describes the design of a software package meant to address all 
of these concerns. It provides a simple way to 
coordinate processing and storage of large datasets by integrating grids and small clusters.

\begin{figure*}
\begin{center}
\includegraphics[height=12.0cm,keepaspectratio]{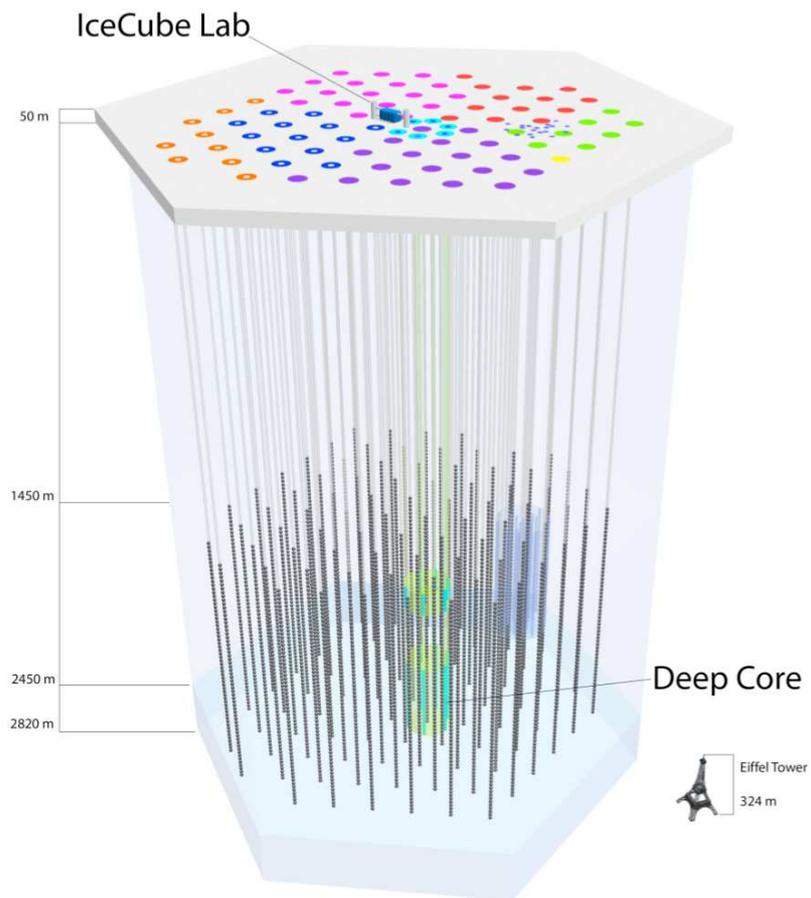}
\end{center}
\caption{The IceCube detector: the dotted lines at the bottom represent
	the instrumented portion of the ice. The circles on the top surface represent
	\emph{IceTop}, a surface air-shower subdetector.}
\label{fig:icecube} 
\end{figure*}
\subsection{The IceCube Detector}
\label{sec:icecube}

The IceCube detector shown in Figure \ref{fig:icecube} is located at the geographic South Pole and was completed at the end of
2010 \cite{icecube:halzen,PhysRevD.87.062002}. It consists of 5160 optical
sensors buried between 1450 and 2450 meters below the surface of the
South Pole ice sheet and is designed to detect interactions of
neutrinos of astrophysical origin \cite{icecube:halzen}. However, it is also sensitive to
downward-going highly energetic muons and neutrinos produced in
cosmic-ray-induced air showers. 
IceCube records ${\sim}10^{10}$ cosmic-ray events per
year. The cos\-mic-ray-in\-duced muons outnumber neutrino-induced
events (including ones from atmospheric origin) by about
500,000:1. They represent a background for most
IceCube analyses and are filtered prior to transfer to the data processing
center in the Northern Hemisphere. Filtering at the data
collection source is required  because of bandwidth limitations on the satellite connection between the detector and the 
processing location \cite{icecube:halzen-klein}. 
About 100 GB of data from the IceCube detector is transferred to the main
data storage facility daily. 
In order to facilitate record keeping, the data is divided into runs, and each run is 
further subdivided into multiple files. The size of each file is
dictated by what is considered optimal for storage and access. Each run
typically consists of hundreds of files, resulting in ${\sim}$400,000 files
for each year of detector operation. 
Once the data has
been transferred, additional, more computationally-intensive event reconstructions are
performed and the data is filtered to select events for various
analyses.  The computing requirements for the various levels of data processing are shown
in Table \ref{tab:proc-needs}.  
In order to develop event reconstructions, perform analyses, and understand systematic
uncertainties, physicists require statistics from Monte Carlo
simulations that are comparable to the data collected by
the detector. This requires thousands of years of CPU
processing time as can be seen from Table \ref{tab:sim-needs}.

\begin{table}[htb]  
\caption[Data processing CPU]{Data processing demands.
Data is filtered on 400 cores at the South Pole using loose selection criteria to
	reduce volume by a factor of
	10 before satellite transfer to the Northern Hemisphere (Level1).
	Once in the North, more computationally intensive event reconstructions are performed
in order to further reduce background contamination (Level2). Further
event selections are made for each analysis channel (Level3). Each run
is equivalent to approximately eight hours of detector livetime and the
processing time is based on a 2.8 GHz core. }
\label{tab:proc-needs}
\centering 
\begin{tabular}{l|r|r}\hline \hline
Filter 			&  Processing time/run & Total per year\\ \hline
Level1 			&  2400 h	& $2.6 \times 10^6$ h \\ 
Level2 			&  9500 h	& $1.0 \times 10^7$ h \\ 
Level3  	    		&  15 h	& $1.6 \times 10^4$ h \\ \hline
\end{tabular}
\end{table}

\begin{table} [htb]   
\caption[Mean runtime of various Monte Carlo simulations.] {Runtime of
	various Monte Carlo simulations of background cosmic-ray shower events and neutrino
	signal with different energy distributions. The median energy is
	based on the distribution of events that trigger the detector. The number of events reflects the typical per-year requirements for IceCube analyses.}
\label{tab:sim-needs}
\centering 
\begin{tabular}{l|c|r|l}\hline \hline
Simulation & Med. Energy\footnotemark  & t/event & events \\ \hline
Air showers & $1.2 \times 10^4$ GeV & 5  ms & ${\sim}10^{14}$ \\ 
Neutrinos & $3.9 \times 10^6$ GeV    & 316 ms  & ${\sim}10^8$ \\ 
Neutrinos &  $8.1 \times 10^1$ GeV   & 53 ms &   ${\sim}10^9$ \\ \hline
\end{tabular}
\end{table}
\footnotetext{1 GeV = $10^{9}$ electronvolts (unit of energy)}

\subsection{IceCube Computing Resources}
\label{sec:iccomp}
The IceCube collaboration is comprised of 43 research institutions
from Europe, North America, Japan, Australia, and New Zealand. Members of the collaboration
have access to 25 different computing clusters and grids in Europe, Japan, Canada and the
U.S. These range from small computer farms of 30 nodes to large grids,
such as the European Grid Infrastructure (EGI), 
Swedish Grid Initiative (SweGrid), Canada's WestGrid and the Open
Science Grid (OSG), that may each have thousands of computing nodes. 
The total number of nodes available to IceCube member
institutions varies with time since much of our use is opportunistic and availability depends on
the usage by other projects and experiments. In total, IceCube simulation has
run on more than 11,000 distinct multicore computing nodes. 
On average, IceCube simulation production has run concurrently on
${\sim}4,000$ cores at any given time since deployment, and it is anticipated to run on
${\sim}5,000$ cores simultaneously during upcoming productions.

\section{IceProd}
\label{ch:statement}  
         The IceProd framework is a software package developed for IceCube
	 with the goal of managing productions across distributed systems
	 and pooling together isolated computing resources that are
	 scattered across member institutions of the Collaboration and beyond.  
	 It consists of a central database and a set of daemons that are
	 responsible for the management of grid jobs and data handling
	 through the use of existing grid technology and network protocols.

     	IceProd makes job scripting easier and sharing productions more efficient. 
     	In many ways it is similar to PANDA Grid, the analysis framework for the PANDA experiment
     	\cite{protopopescu2011panda}, in that both tools are
     	distributed systems based on a central database and an interface to
     	local batch systems. Unlike PANDA Grid which depends heavily on AliEn, the grid
     	middleware for the ALICE experiment \cite{alien}, and on the ROOT analysis framework \cite{ROOT},
    	IceProd was built in-house with minimal software requirements and is not dependent
     	on any particular middleware or analysis framework. It is designed
     	to run completely in user space with no administrative access, allowing greater flexibility in
     	installation. IceProd also includes a built-in monitoring system with no dependencies 
     	on any external tools for this purpose. These properties make IceProd a very
    	lightweight yet powerful tool and give it a greater scope beyond IceCube-specific applications. 

     The software package includes a
     set of libraries, executables and daemons that communicate with the
     central database and coordinate to share responsibility for the
     completion of tasks. The details of job submission and management
	 in different grid environments are
     abstracted through the use of plugin modules that will be discussed in Section \ref{sec:plugins}.  

	 IceProd can be used to
     integrate an arbitrary number of sites including clusters and
	 grids. It is, however, not a replacement for other cluster and grid
	 management tools 
	 or any other middleware. 	 Instead, it runs on top of these as a separate layer
	 providing additional functionality. IceProd fills a gap between the user or production manager
	 and the powerful middleware and batch system
	 tools available on computing clusters and grids.

	 Many of the existing middleware tools, including Con\-dor-C, Globus and CREAM,  
	 make it possible to interface any number of computing clusters 
	 into a larger pool. However, most of these tools need to be installed and configured by system
	 administrators and, in some cases, customization for general purpose applications is not feasible.
	 In contrast to most of these applications, IceProd runs at
	 the user level and does not require administrator privileges. This makes
	 it possible for individual users to build large production systems by
	 pooling small computational resources together.  

	 Security and data
     integrity are concerns in any software architecture that
     depends heavily on communication through the Internet.
     IceProd includes features aimed at minimizing security and
     data corruption risks. Security and data integrity are addressed in
	 Section \ref{sec:security}.

	 The IceProd client provides a graphical user interface (GUI) for 
	 configuring simulations and submitting jobs through a ``production
	 server.''  
	 It provides a method for recording all the software versions,
	 physics parameters, system settings,
        and other steering parameters associated with a job in a central production database. IceProd
        also includes a web interface for visualization and live monitoring of
	 datasets. Details about the GUI client and a text-based client are
	 discussed in Section \ref{sec:client}.

\section{Design Elements of IceProd}
\label{sec:design}

	The IceProd software package can be logically divided into the
	following components or software libraries:
	\begin{itemize}
	\item \emph{iceprod-core}\textemdash a set of modules and libraries of common
	use throughout IceProd.
	\item \emph{iceprod-server}\textemdash a collection of daemons and libraries to manage and schedule job 
	submission and monitoring.
	\item \emph{iceprod-modules}\textemdash a collection of predefined classes that provide an 
	interface between IceProd and an arbitrary task to be performed on
	a computing node, as defined in Section \ref{iceprodmodules}.
	\item \emph{iceprod-client}\textemdash a client (both graphical and text)
	that can download, edit, and submit dataset steering files to be processed.
	\item A database that stores configured parameters, libraries (including version information), 
	job information, and performance statistics.
	\item A web application for monitoring and controlling dataset processing. 
	\end{itemize}
	These components are described in further detail in the following sections.

\subsection{IceProd Core Package}
	The \emph{iceprod-core} package contains modules and libraries common to all other IceProd packages. 
	These include classes and methods for writing and parsing XML files and transporting data. The classes 
	that define job execution on a host are contained in this package.
	The \emph{iceprod-core} also includes an interpreter (Section \ref{sec:expressions})
	for a simple scripting language that provides some flexibility for parsing XML steering files.

\subsubsection{The JEP}
		One of the complications of operating on heterogeneous systems is
		the diversity of architectures, operating systems, and compilers.
		IceProd uses HTCondor's NMI-Metronome build and
		test system \cite{condor:nmi} for building the IceCube software
		on a variety of platforms and storing the built packages on a
		server. As part of the management of each job, IceProd submits a Job Execution Pilot
		(JEP) to the cluster/grid queue. This script determines what
		platform a job is running on and, after contacting the monitoring server, 
		which software package
		to download and execute.  During runtime, the JEP 
		performs status updates through the monitoring server via remote
		procedure calls using XML-RPC \cite{winer99}.
		This information is updated on the database and is displayed on the
		monitoring web interface. Upon completion, the JEP removes
		temporary files and directories created for the job. 
		Depending on the configuration, it will also cache a copy of the software used, making it 
		available for future JEPs. 
		When caching is enabled, an MD5 checksum is performed on the cached software and
		compared to what is stored on the server in order to avoid using
		corrupted or outdated software. 
		
\begin{figure}
\begin{center}
\includegraphics[width=8.0cm,keepaspectratio]{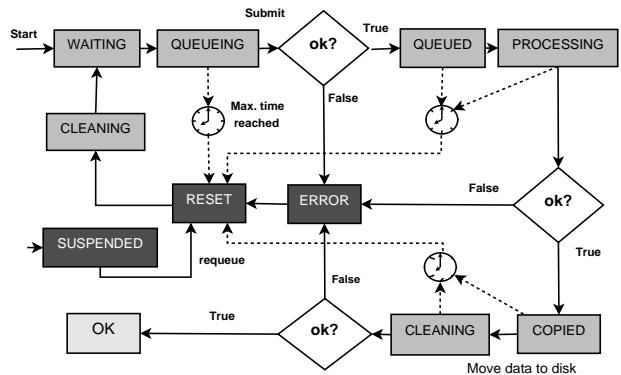}
\end{center}
\caption{State diagram for the JEP.  Each of the nonerror states through which a job passes includes a
            configurable timeout. The purpose of this timeout is to account for
            any communication errors that may have prevented a job from setting
            its status correctly.}
\label{fig:jobstate} 
\end{figure}

		Jobs can fail under many circumstances. These failures include failed submissions due to
		transient system problems and execution failures due to problems
		with the execution host. At a higher level, errors specific to
		IceProd include communication problems with the monitoring daemon or
		the data repository. In order to account for possible transient
		errors, the design of IceProd includes a set of states through which
		a job will transition in order to guarantee successful completion
		of a well-configured job. The state diagram for an IceProd job is
		depicted in Figure \ref{fig:jobstate}.

\subsubsection{XML Job Description}
\label{sec:xml}
		In the context of this document, a dataset is defined as a
		collection of jobs that share a basic set of scripts and software
		but whose input parameters depend on the ID of each individual job. 
		A configuration or steering file describes the tasks to be
		executed for an entire dataset. IceProd steering files are XML
		documents with a defined schema. These steering files include information
		about the specific software versions used for each of the sections,
		known as trays (a term borrowed from IceTray, the C++ software framework used
		by the IceCube Collaboration \cite{icecube:deyoung}). 
		An IceProd tray represents an instance of an environment
		corresponding to a set of libraries and executables and a chain
		of configurable modules with corresponding parameters and input
		files needed for the job.
		In addition, there is a header section for user-defined parameters and
		expressions that are globally accessible by different modules.

\subsubsection{IceProd XML expressions}
\label{sec:expressions}
		A limited programming language was developed in order to allow
		more scripting flexibility that depends on runtime parameters
		such as job ID, run ID, and dataset ID. This lightweight, embedded, domain-specific language (DSL) 
		allows for a single XML
		job description to be applied to an entire dataset following an
		SPMD (single process, multiple data) paradigm. It is powerful
		enough to give some flexibility but sufficiently restrictive to
		limit abuse. Examples of
		valid expressions include the following:
		
		\begin{itemize}
		\item \texttt{\$args(<var>)}\textemdash  a command line argument passed to the job (such as job ID or dataset ID).
		\item \texttt{\$steering(<var>)}\textemdash a  user defined variable.
		\item \texttt{\$system(<var>)}\textemdash a system-specific parameter defined by the server.
		\item \texttt{\$eval(<expr>)}\textemdash a mathematical or logical expression (in Python).
		\item \texttt{\$sprintf(<format>,<list>)}\textemdash string formatting.
		\item \texttt{\$choice(<list>)}\textemdash random choice of an element from the list.
		\end{itemize}

        The evaluation of such expressions is recursive and allows for some
        complexity. However, there are limitations in place that
        prevent abuse of this feature. As an example, \texttt{\$eval()}
        statements prohibit such things as loops and import statements that
        would allow the user to write an entire program within an
        expression. There is also a limit on the number of recursions in order
        to prevent closed loops in recursive statements.

\begin{figure*}[!ht]
\begin{center}
\includegraphics[width=11.0cm,keepaspectratio]{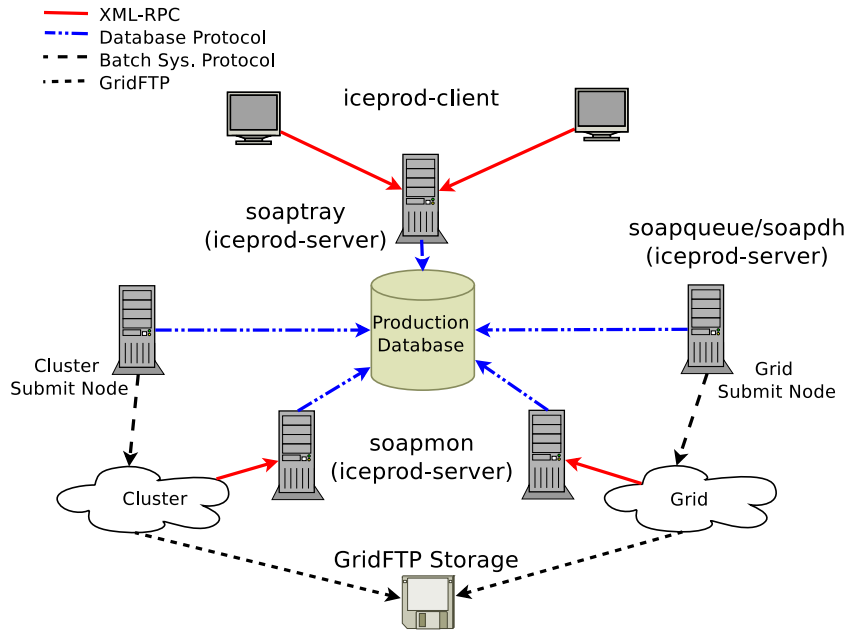}
\end{center}
\caption{Network diagram of IceProd system. The IceProd clients and JEPs communicate 
	    with \emph{iceprod-server} modules via XML-RPC. Database calls are 
		restricted to \emph{iceprod-server} modules. Queueing daemons
		called \emph{soapqueue}
		are installed at each site and periodically query the database for
		pending job requests. The \emph{soapmon} server receives
		monitoring update from the jobs. An instance of \emph{soapdh}
		handles garbage collection and any post processing tasks after job completion.
}
\label{fig:iceprod-network} 
\end{figure*}

\subsection{IceProd Server}
\label{sec:server}
\begin{figure}[!ht]
\begin{center}
\includegraphics[width=7.6cm,keepaspectratio]{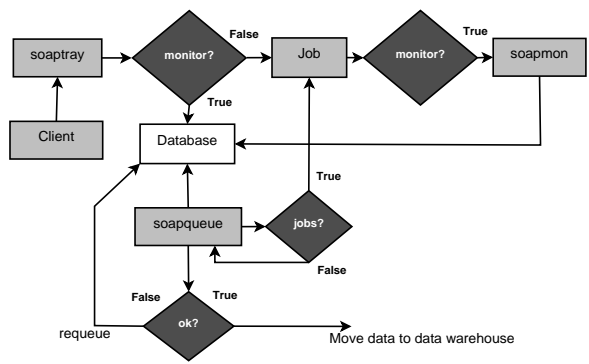}
\end{center}
\caption{State diagram of queuing algorithm. The iceprod-client sends
	requests to the \emph{soaptray} server which then loads the information to
	the database (in production mode) or directly submits jobs to the
	cluster (in unmonitored mode). The \emph{soapqueue} daemons periodically query the database for
	pending requests and handle job submission in the local cluster.}
\label{fig:iceprod-submit} 
\end{figure}

\begin{figure*}[!ht]
\begin{center}
\includegraphics[width=14.0cm,height=6.5cm]{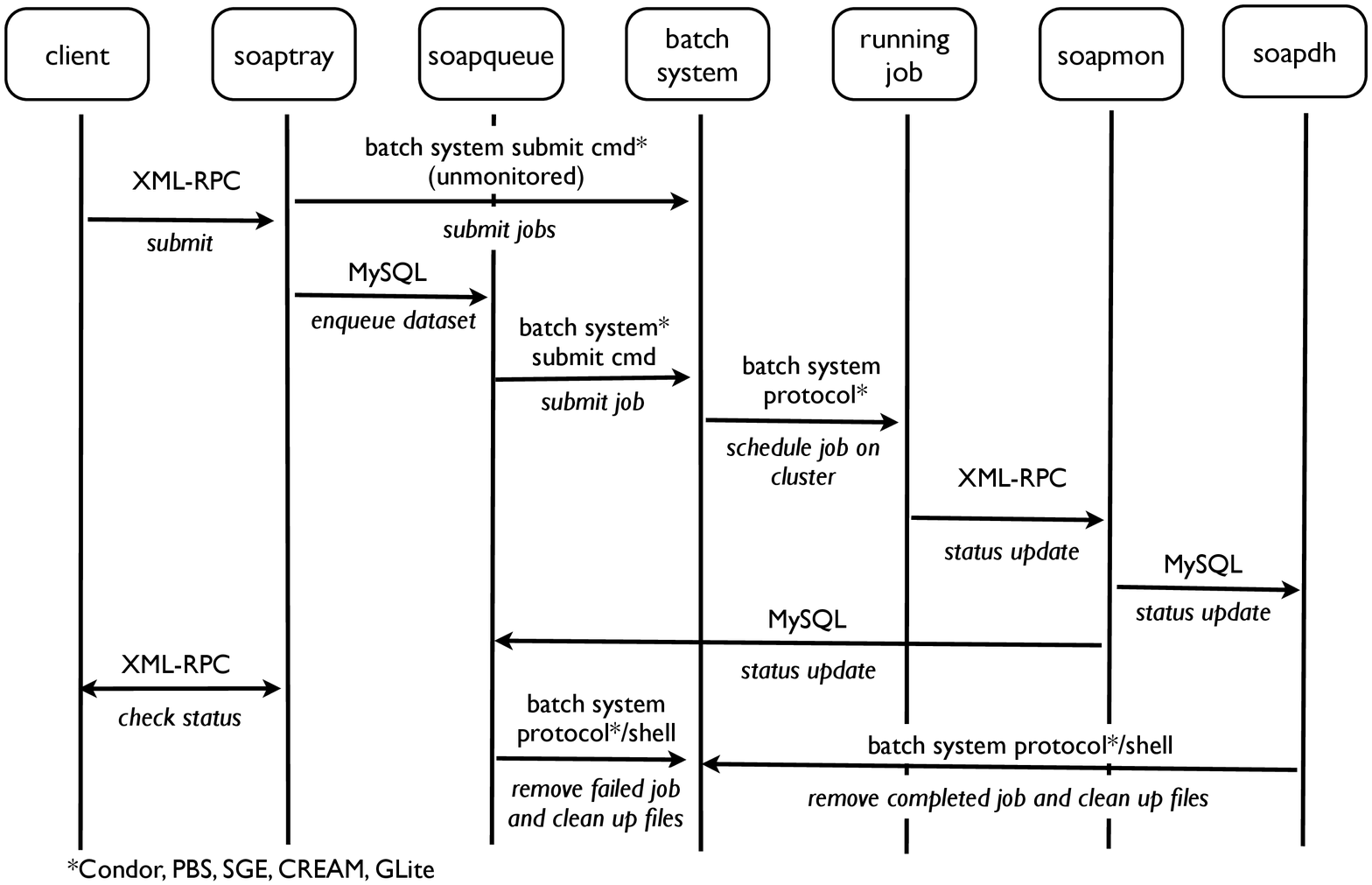}
\end{center}
\caption{Data flow for job submission, monitoring and removal.
	Communication between server instances (labeled ``soap*'') is handled through a
database. Client/server communication and monitoring updates are handled
via XML-RPC. Interaction with the grid or cluster is handled through a
set of plugin modules and depends on the specifics of the system.}
\label{fig:swim} 
\end{figure*}

		The \emph{iceprod-server} package is comprised of four daemons and their respective libraries: 
        \begin{enumerate} 
			\item \emph{soaptray}\footnote{The prefix \emph{soap} is
				used for historical reasons. The original implementation
					of IceProd relied on SOAP for remote 
					procedure calls. This was replaced by XML-RPC which has better support in Python.}\textemdash an HTTP server that receives client
			XML-RPC requests for scheduling jobs and steering
			information which then uploaded to the database. 
			\item \emph{soapqueue}\textemdash a daemon that queries the database
			for available tasks to be submitted to a particular cluster
			or grid. This daemon is also responsible for submitting jobs
			to the cluster or grid through a set of plugin classes.
			\item \emph{soapmon}\textemdash a monitoring HTTP server that
			receives XML-RPC updates from jobs during execution and performs status 
			updates to the database.  
			\item \emph{soapdh}\textemdash a data handling/garbage collection
			daemon that removes temporary files and performs any postprocessing tasks.
        \end{enumerate}
		There are two modes of operation. The first is an unmonitored mode in which
        jobs are simply sent to the queue of a particular system.
        This mode provides a tool for scheduling jobs that don't need to be recorded and does not require a database.
        In the second mode, all parameters are stored in a
        database that also tracks the progress of
        each job. The \emph{soapqueue} daemon running at each of the participating
        sites periodically queries the database to check if any tasks have been
        assigned to it. It then downloads the steering configuration and submits
        a given number of jobs to the cluster or grid where it is running. 
		The number of jobs that IceProd maintains in the queue at
		each site can be configured individually according to the
		specifics of each cluster, including the size of the cluster and local queuing
        policies. 
        \mbox{Figure \ref{fig:iceprod-network}} is a graphical representation
		that describes the interrelation of these daemons. 
		The state diagram in \mbox{Figure \ref{fig:iceprod-submit}} illustrates the role of the daemons in
		dataset submission while 
		\mbox{Figure \ref{fig:swim}} illustrates the flow of information through the various protocols.

\subsubsection{IceProd Server Plugins}
\label{sec:plugins}

        In order to abstract the process of job submission from the framework for the various types
        of systems, IceProd defines a Grid base class that provides an
        interface for queuing jobs. The Grid base class interface includes a set
		of methods for queuing and removing jobs, performing status checks, and setting attributes such as
        job priority and maximum allowed wall time and job requirements such as disk space and memory usage. 
        The set of methods defined by this base class include but
		are not limited to:
		\begin{itemize}
		\item {\bf WriteConfig}: write protocol-specific submission
		scripts (i.e., a JDL job description file in the case of CREAM or
				gLite or a shell script with the appropriate PBS/SGE headers).
		\item {\bf Submit}: submit jobs and record the job ID in the local queue.
		\item {\bf CheckJobStatus}: query job status from the queue.
		\item {\bf Remove}: cancel/abort a job.
		\item {\bf CleanQ}: remove any orphan jobs that might be left in the queue.
		\end{itemize}
		The actual implementation of these methods is done by a set of
		plugin subclasses that launch the corresponding commands or
		library calls, as the case may be. In the case of PBS and SGE,
		most of these methods result in the appropriate system calls to
		\emph{qsub}, \emph{qstat}, \emph{qdel}, etc. For other systems,
		these can be direct library calls through a Python API.
		IceProd contains a growing library of plugins, including classes for interfacing with batch systems such as
        HTCondor, PBS and SGE as well as grid systems like Globus, gLite, EDG, CREAM and ARC. In addition, one can easily implement user-defined
        plugins for any new type of system that is not included in this list.

\subsection{IceProd Modules}
\label{iceprodmodules}
		The \emph{iceprod-modules} package is a collection of configurable modules with a common interface.
		These represent the atomic tasks to be
		performed as part of the job. They are derived from
		a base class \emph{IPModule} and provide a standard interface that
		allows for an arbitrary set of parameters to be configured in the
		XML document and passed from the IceProd framework. In turn, the
		module returns a set of statistics in the form of a string-to-float
		dictionary back to the framework so that it can be 
		recorded in the database and displayed on the monitoring web interface.
		By default, the base class will report the module's CPU usage,
		but the user can define any set of values to be reported, such as
		number of events that pass a given processing filter. IceProd
		also includes a library of predefined modules for performing common
		tasks such as file transfers through GridFTP, tarball manipulation,
		etc.

\subsection{External IceProd Modules}
		Included in the library of predefined modules is a special module
        that has two parameters:
		\emph{class} and \emph{URL}. The first is a string
		that defines the name of an external IceProd module and the second
		specifies a URL for a (preferably version-controlled)
		repository  where the external module code
		can be found. 
		Any other parameters passed to this module are assumed
		to belong to the referred external module and will be ignored.
		This allows for the use of user-defined modules without
		the need to install them at each IceProd site. External modules
		share the same interface as any other IceProd module. 
		External modules are retrieved and cached by the server at the time of submission. 
		These modules are then included as file dependencies for the jobs, thus preventing the need 
		for jobs to directly access the file code repository. 
		Additional precautions, such as enforcing the use of secure protocols for URLs, must be taken to avoid security risks.

\subsection{IceProd Client}
\begin{figure*}[!ht]
\begin{center}
\includegraphics[width=15.0cm,height=14.0cm,keepaspectratio]{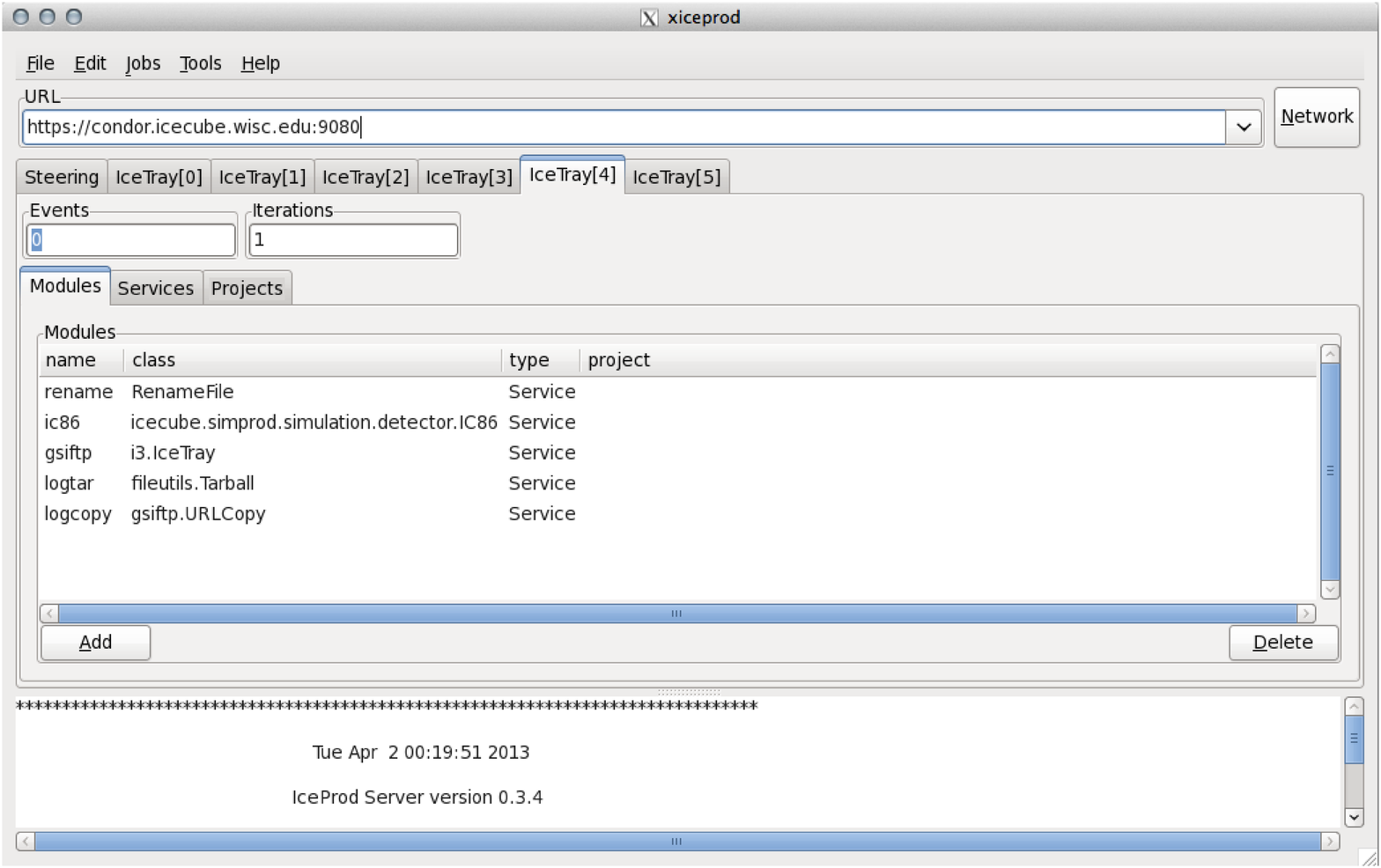}
\end{center}
\caption[The xiceprod client]{The \emph{iceprod-client} uses pyGtk and
	provides a graphical user interface to IceProd. It is both a
		graphical editor of XML steering files and an XML-RPC client for
		dataset submission.}
\label{fig:xiceprod} 
\end{figure*}
\label{sec:client}
		The \emph{iceprod-client} package contains two applications for interacting with
		the server and submitting datasets. One is a PyGTK-based GUI
		(see Figure \ref{fig:xiceprod}) and the
		other is a text-based application that can run as a command-line
		executable or as an interactive shell. Both of these applications
		allow the user to download, edit, and submit steering configuration files as well
	        as control datasets running on the IceProd-controlled grid. The graphical interface
		includes drag and drop features for moving modules around and
		provides the user with a list of valid parameters for known modules.
		Information about parameters for external modules is not included
		since these are not known a priori. The interactive shell also
		allows the user to perform grid management tasks such as starting
		and stopping a remote server and adding and removing production sites participating in 
		the processing of a dataset. The user can also perform job-specific actions such as suspension and resetting of jobs.

%
%
\subsection{Database}
	At the time of this writing, the current implementation of IceProd works exclusively with a MySQL 
	database, but all database calls are handled by a database module that abstracts queries 
	from the framework and could be easily replaced by a different relational database. This section describes the 
	relational structure of the IceProd database.

\label{DBStructure}
	Each dataset is defined by a set of modules and parameters that operate on separate
	data (single process, multiple data). At the top level of the database structure is the dataset table. 
	The \textit{dataset ID} is the unique identifier for each dataset, though it is possible 
	to assign a mnemonic string alias. The tables in the IceProd database are logically divided into two distinct classes that could in 
	principle be entirely different databases. The first describes a steering file or dataset configuration 
	(items \ref{db:dataset}\textendash \ref{db:cparameter} and \ref{db:taskrel} in the list below) and the second is a job-monitoring 
	database (items \ref{db:job} and \ref{db:task}). The most important tables are described below.

	\begin{enumerate}
	\item {\bf dataset}\label{db:dataset}: contains a unique identifier as well as attributes to describe and categorize the dataset, including a textual description. 
	\item {\bf steering-parameter}\label{db:steerparam}: describes general global variables that can be referenced from any module.
	\item {\bf meta-project}: describes a software environment including libraries and executables.
	\item {\bf tray}: describes a grouping of modules that will execute given the same software environment or \emph{metaproject}.
	\item {\bf module}: specifies an instance of an IceProd Module class.
	\item {\bf cparameter}\label{db:cparameter}: contains all the configured parameters associated with a module.
	\item {\bf job}\label{db:job}: describes each job in the queue related to a dataset, including the state and host where the job is executed. 
	\item {\bf task}\label{db:task}: keeps track of the state of a task in a way similar to what is done in the jobs table. 
	A task represents a subprocess for a job in a process workflow. More details on this will be provided in Section \ref{sec:dags}.
	\item {\bf task-rel}\label{db:taskrel}: describes the hierarchical relationship between tasks.
	\end{enumerate}

%
%
\subsection{Monitoring}
       The status updates and statistics are reported by the JEP via XML-RPC
	   to \emph{soapmon} and stored in the database, and provide useful information for
       monitoring the progress of processing datasets and for
       detecting errors. The updates include status changes
       and information about the execution host as well as job
       statistics. This is a multi-threaded server that can run as a
       stand-alone daemon or as a CGI script within a more robust
       web server. The data collected from each job are made available for analysis, and
       patterns can be detected with the aid of visualization tools as
       described in the following section.

\subsubsection{Web Interface}
\begin{figure*}[!ht]
\begin{center}
\includegraphics[width=14.0cm,keepaspectratio]{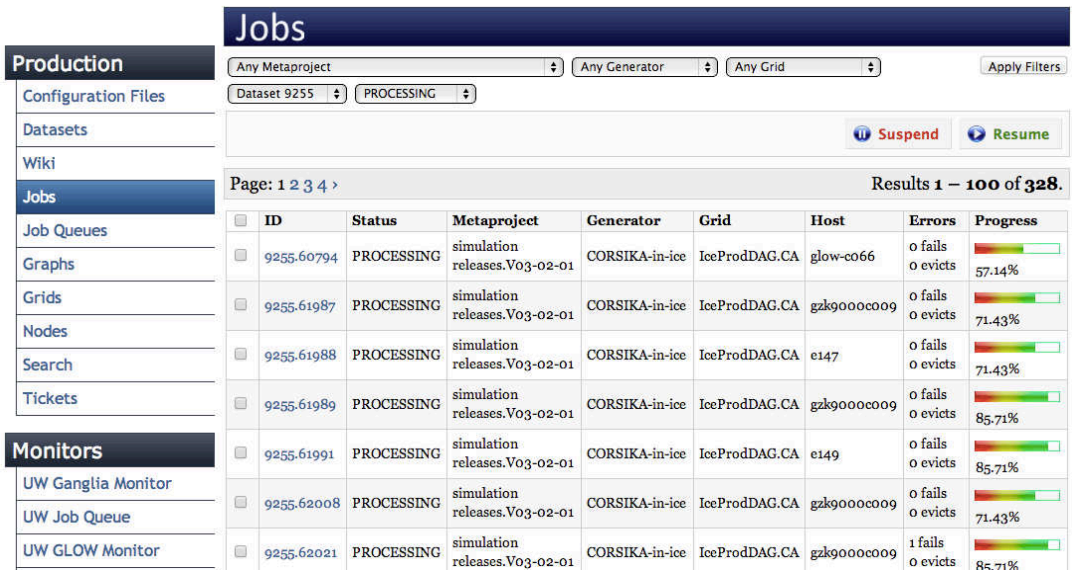}
\end{center}
\caption{A screen capture of the web interface that allows the monitoring of
	ongoing jobs and datasets. The monitoring web interface has a number of
		views with different levels of detail. The view shown displays
		the job progress for active jobs within a dataset. The web
		interface provides authenticated users with buttons to control
		datasets and individual jobs.}
\label{fig:web} 
\end{figure*}  
       The current web interface for IceProd was designed to work independently of the
       IceProd framework but utilizes the same database. It is
       written in PHP and makes use of the CodeIgniter framework \cite{codeigniter}. Each of the simulation and 
       data-processing web-monitoring tools provide different views, which
       include, from top level downward: 

		\begin{itemize}
		\item general view: displays all datasets filtered by status, type, grid, etc.
		\item grid view: shows all datasets running on a particular site.
		\item dataset view: displays all jobs and accompanying statistics for a given dataset, including every site that it is running on.
		\item job view: shows each individual job, including the status, job statistics, execution host, and possible errors.
		\end{itemize}
		There are some additional views that are applicable only to the processing of real IceCube detector data:
		\begin{itemize}
		\item calendar view: displays a calendar with a color coding that indicates the status of jobs associated with data taken on a particular date.
		\item day view: shows the status of jobs associated with a given
		calendar day of data taking. 
		\item run view: displays the status of jobs associated with a particular detector run.
		\end{itemize}

		The web interface also provides the functionality to control jobs and datasets by authenticated users.
		This is done by sending commands to the \emph{soaptray} daemon using the XML-RPC protocol. Other features of the interface include
		graphs displaying completion rates, errors and number of jobs in
		various states. Figure \ref{fig:web} shows a screen capture of
		one of a number of views from the web interface.

\subsubsection{Statistical Data}
		One aspect of IceProd that is not found in most grid middleware is the built-in collection of 
		user-defined statistical data. Each IPModule instance is passed a
		string-to-float dictionary to which the JEP
		can add entries or increment a given value. IceProd collects these data in the central database
		and displays them on the monitoring page. Statistics are reported individually for each job and collectively for the whole dataset 
		as a sum, average and standard deviation. The typical types of information collected on IceCube jobs include 
		CPU usage, number of  events meeting predefined physics criteria, and number of calls to a particular module.

%
%
\subsection{Security and Data Integrity}
\label{sec:security}
    When dealing with network applications, one must always be concerned with security and data 
	integrity in order to avoid compromising privacy and the validity of scientific results.
    Some effort has been made to minimize security risks in the design and implementation of IceProd. 
	This section will summarize the most significant of these. Figure \ref{fig:iceprod-network} 
  shows the various types of network communication between the client, server, and worker node.
    
\subsubsection{Authentication}
	Authentication in IceProd can be handled in two ways:
	IceProd can authenticate dataset submission against an LDAP server
	or, if one is not available, authentication is handled by  means of direct database authentication.  
	LDAP authentication allows the IceProd administrator to restrict usage to 
	individual users that are responsible for job submissions and are accountable for improper use so direct database 
	authentication should be disabled whenever LDAP is available.
	This setup also precludes the need to distribute database passwords and thus prevents users from being able to directly query the database via a MySQL client.
	
	When dealing with databases, one also needs to be concerned about
	allowing direct access to the database and passing login credentials
	to jobs running on remote sites.  For this reason, 
	all monitoring calls are done via XML-RPC, and the only direct queries are performed by the server, which typically 
	operates behind a firewall on a trusted system. The current web
	interface does make direct queries to the database;
	a dedicated read-only account is used for this purpose.

 \subsubsection{Encryption}                   
	Both \emph{soaptray} and \emph{soapmon} can be configured to
	use SSL certificates in order to encrypt all data communication between client and server. 
	The encryption is done by the HTTPS server with either a self-signed
	certificate or, preferably, with a 
	certificate signed by a trusted Certificate Authority (CA).
	This is recommended for client-server communication for \emph{soaptray} but is generally not considered necessary for
	monitoring information sent to \emph{soapmon} by the JEP as this is not considered sensitive enough to justify 
	the additional system CPU resources required for encryption.

 \subsubsection{Data Integrity}   
\label{ch:integrity}
	In order to guarantee data integrity, an MD5 checksum or digest is generated for each file
	that is transmitted. This information is stored in the database and is checked against the file after
	transfer. IceProd data transfers support several protocols, but the preference is to
	rely primarily on GridFTP, which makes use of GSI authentication \cite{gridftp,globus:grid-security}.                   
	        
	An additional security measure is the use of a temporary passkey that is assigned
	to each job at the time of submission. This passkey is used for authenticating communication between the job and the
	monitoring server and is only valid during the duration of the job. If the job is reset, this passkey will be changed
	before a new job is submitted. This prevents stale jobs that might be left running from making monitoring updates after
	the job has been reassigned.

%
%
\section{Intrajob Parallelism}
\label{sec:dags}
	As described in Section~\ref{sec:xml}, a single IceProd job consists of a number of trays and modules
	that execute different parts of the job, for example, a simulation chain.
	These trays and modules describe a workflow with a set of interdependencies, where the output from
	some modules and trays is used as input to others.
	Initial versions of IceProd ran jobs solely as monolithic scripts that executed these modules serially
	on a single machine.
	This approach was not very efficient because it did not take advantage of the workflow structure implicit
	in the job description.
	
\begin{figure}[h]
\begin{center}
\includegraphics[width=4.5cm,keepaspectratio]{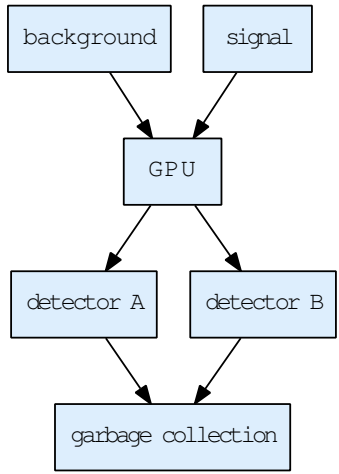}
\end{center}
\caption{A simple DAG in IceProd. This DAG corresponds to a typical
	IceCube simulation. The two root vertices require standard computing
	hardware and produce different types of signal. Their output is 
	then combined and processed on GPUs. The output is then 
	used as input for two different detector simulations.}
\label{fig:dag1} 
\end{figure}

\begin{figure*}
\begin{center}
\includegraphics[width=15.0cm,keepaspectratio]{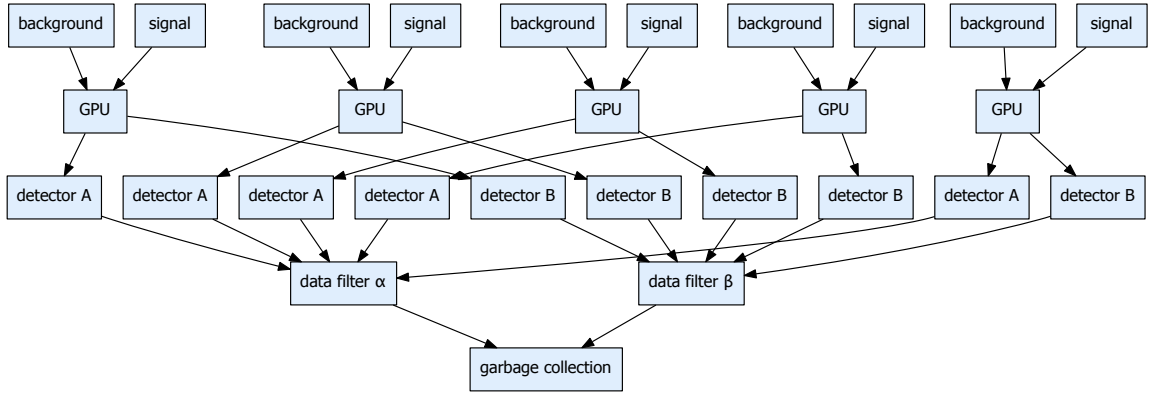}
\end{center}
\caption{A more complicated DAG in IceProd with multiple inputs and
	multiple outputs that are eventually merged into a single output.
		The vertices in the second level run on computing nodes equipped with GPUs.}
\label{fig:dag2} 
\end{figure*}

	To address this issue, IceProd includes a representation of a job as a directed, acyclic graph (DAG)
	of tasks.
	Jobs are recharacterized as groups of arbitrary tasks and modules that are defined by users in a job's XML steering file,
	and each task can depend on any number of other tasks in the job.
	This workflow is encoded in a DAG, where each vertex represents a single instance of a task to be executed
	on a computing node, and edges in the graph indicate dependencies between tasks (see Figures~\ref{fig:dag1}
	and \ref{fig:dag2}).
	DAG jobs on the cluster are executed by means of the HTCondor DAGMan
	which is a workflow manager developed by the HTCondor group at the University of Wisconsin\textendash Madison
	and included with the HTCondor batch system~\cite{condor:dagman}. 	
	
	
	For IceCube simulation production, IceProd has utilized the DAG
	support in two specific cases: improving task-level parallelism and
	running jobs that utilize graphics processing units (GPUs) for
	portions of their processing.

\subsection{Task-level Parallelism}
	In addition to problems caused by coarse-grained requirements specifications, monolithic jobs also underutilize
	cluster resources.
	As shown in Figure~\ref{fig:dag1}, portions of the workflow within a job are independent; however, if a job is
	monolithic, these portions will be run serially instead of in parallel.
	Therefore, although the entire simulation can be parallelized by submitting multiple jobs to different machines,
	this opportunity for additional parallelism is not exploited by monolithic jobs.
	
	Support for breaking a job into discrete tasks is now included in the HTCondor IceProd plugin as described above, and
	similar features have been developed for the PBS and Sun Grid Engine plugins.
	This enables faster execution of individual jobs by utilizing more computing nodes; however, one limitation of
	this implementation is that DAG jobs are restricted to a specific type of cluster, and DAG jobs cannot distribute
	tasks across multiple sites. 

\subsection{DAGs Based on System Requirements}
	Individual parts of a job may have different system hardware and
	software requirements. Breaking these up into tasks that run on
	separate nodes allows for better utilization of resources.
	The IceCube detector
	simulation chain is a good example of this scenario in which
	tasks are distributed across
	computing nodes with different hardware resources.
	
	Light propagation in the instrumented volume of ice at the South
	Pole is difficult to model, but recent developments in IceCube's simulation include a much faster approach for simulating direct 
	propagation of photons in the optically complex Antarctic ice
	\cite{Aartsen:2013rt,icecube:ice} by using general-purpose GPUs.  
	This new simulation module is much faster than a CPU-based implementation and more accurate than using 
	parametrization tables~\cite{Chirkin2013141}, but the rest of the simulation requires standard CPUs.
	When executing an IceProd job monolithically, only one set of cluster requirements can be applied when it is
	submitted to the cluster.
	Accordingly, if any part of the job requires use of a GPU, the entire monolithic job must be scheduled on a
	cluster machine with the appropriate hardware.
	
	As of this writing, IceCube has the potential to access ${\sim}20,000$ CPU cores distributed throughout the world, but only 
	a small number of these nodes are equipped with GPU cards.
	Because the simulation is primarily CPU bound, the pool of GPU-equipped nodes is not sufficient to run all
	simulation jobs in an acceptable amount of time.
	Additionally, this would be an inefficient use of resources, since executing the CPU-oriented portions of
	monolithic jobs would leave the GPU idle for periods of time. 
	In order to solve this problem, the modular design of the IceCube simulation design is used to divide the
	CPU- and GPU-oriented portions of jobs into separate tasks in a DAG.
	Since each task in a DAG is submitted separately to the cluster, their requirements can be specified independently
	and CPU-oriented tasks can be executed on general-purpose grid nodes while photon propagation tasks can be
	executed on GPU-enabled machines, as depicted in Figure~\ref{fig:dag2}.

%
%
\section{Applications}
    IceProd's highly configurable nature lets it serve the needs of
    many different applications, both inside and beyond the IceCube Collaboration.

%
%
\subsection{IceCube Simulation Production}
	The IceCube simulations are based on a modular software framework called
	\emph{IceTray} in which modules are executed in sequential order. 
	Data is passed between modules in the form of a ``frame'' object.
	IceCube simulation modules represent different steps in the
	generation and propagation of particles, in-ice light propagation,
	signal detection, and simulation of the electronics and data
	acquisition hardware. These modules are ``chained'' together in a
	single \emph{IceTray} instance but can also be broken into
	separate instances configured to write
	intermediate data files. This allows for breaking up the simulation
	chain into multiple IceProd tasks in order to optimize the use of
	resources as described in Section \ref{sec:dags}. 

	For IceCube, Monte Carlo simulations are the most computationally intensive task,
	 which is dominated by the production of background cosmic-ray showers (see Table
	\ref{tab:sim-needs}). A typical Monte Carlo simulation lasts on the order of 8 hours but corresponds to only four seconds of detector livetime. 
	In order to generate sufficient statistics, IceCube simulation production
	needs to make use of available computing resources which are distributed across the world. Table
	\ref{tab:sites} lists all of the sites that have participated in Monte Carlo production.

\begin{table}[h]
\small
\caption[Sites participating in IceCube Monte Carlo production by country]{Sites participating
	in IceCube Monte Carlo production by country.}
\label{tab:sites}
\centering 
\begin{tabular}{l|l|r}\hline 
Country & Queue Type & No. of Sites \\ \hline \hline
Sweden & ARC & 2  \\ \hline
Canada & PBS & 2 \\ \hline
Germany 	& SGE & 1 \\ 
 		& PBS & 3 \\ 
 		& CREAM & 4 \\ \hline
Belgium 	& PBS & 2 \\ \hline
USA 		& HTCondor & 4 \\ 
 		& PBS & 3 \\ 
		& SGE & 4 \\  \hline
Japan 	& HTCondor & 1\\ \hline
\end{tabular}
\end{table}


%
%
\subsection{Off-line Processing of the IceCube Detector Data}
\label{ch:offline}
        IceProd was designed primarily for managing the production of Monte Carlo simulations for IceCube,
	but it has also been successfully
	adopted for managing the processing and reconstruction of experimental data collected by the
	detector.  
	This data collected by IceCube and previously described in Section \ref{sec:icecube} must undergo
	multiple steps of processing, including calibration, 
	multiple-event track reconstructions, and  sorting into various
	analysis channels based on predefined criteria. IceProd has proved to
	be an ideal framework for processing this large volume of data.
	
	For off-line data processing, the existing features in IceProd are used for job
	submission, monitoring, data transfer, verification, and error
	handling. However, in contrast to a Monte Carlo production dataset where the number of
	jobs are defined a priori, a configuration for off-line processing
	of experimental data initiates with an empty dataset of zero jobs. 
	A separate script is then run over the data in order to map a job to a particular file (or group of files) and to generate
	MD5 checksums for each input file.

    Additional minor modifications were needed in order to support the desired features in off-line processing.
    In addition to the tables described in section \ref{DBStructure}, a
    \emph{run} table was created to keep records of runs and dates associated with each file and 
    unique to the data storage structure. All data collected during a \emph{season} (or a one year cycle)
    are processed as a single IceProd dataset. This is because, for
    each IceCube season, all the data collected is processed with the same set
    of scripts, thus following the SPMD model. A job for such a dataset consists of all the
    tasks needed to complete the processing of a single data file. 

	Off-line processing takes advantage of the IceProd built-in system for
	collecting statistics in order to provide information through web interface about the number of
	events that pass different quality selection criteria from completed
	jobs. 
	Troubleshooting and error correction of
	jobs  during processing is also facilitated by IceProd's real-time feedback system accessible through the
	web interface. The data integrity checks discussed in Section \ref{ch:integrity}
	also provide a convenient way to validate data written to storage and to
	check for errors during the file transfer task.

\subsection{Off-line Event Reconstruction for the HAWC Gamma-Ray Observatory}
\label{hawc}
         IceProd's scope is not limited to IceCube. Its design is general
	 enough to be used for other applications. The
	 High-Altitude Water Cherenkov (HAWC) Observatory \cite{hawc} has recently begun using
	 IceProd for its own off-line event reconstruction and data transfer \cite{hawcprod}. 
	 HAWC has two main computing centers, one located at the University of Maryland and one at UNAM in Mexico City. 
	 Data is collected from the detector in Mexico and then replicated to UMD. The event reconstruction for HAWC is similar in nature to IceCube's 
	 data processing. Unlike IceCube's Monte Carlo production, it is I/O bound and better suited for a local cluster rather than a distributed grid environment.
	 The HAWC Collaboration has made important contributions to the development of IceProd and 
	 maintained active collaboration with the development team.

%
%
\subsection{Deploying an IceProd Site}
    Deployment of an IceProd instance is relatively easy. Installation of the software packages is handled through Python's built-in Module Distribution Utilities package.
    If the intent is to create a stand-alone instance or to start a new grid, 
    the software distribution also includes scripts that define the MySQL tables required for IceProd.
 
    After the software is installed, the server needs to be configured through an INI-style file.
    This configuration file contains three main sections:  general queueing options,
    site-specific system parameters, and job environment. 
    The queueing options are used by the server plugin to help configure submission (e.g. selecting a queue or passing custom directives to the queueing system).   
    System parameters can be used to define the location of a download directory on a shared filesystem
    or a scratch directory to write temporary files. The job environment can be modified by the server configuration to
    modify paths appropriately or set other environment variables.

    If the type of grid/batch system for the new site is already supported, the IceProd instance can be configured
    to use an existing server plugin, with the appropriate local queuing options.
    Otherwise, the server plugin must be written, as described in Section \ref{sec:plugins}.

%
%
\subsection{Extending Functionality}
	The ease of adaptation of the framework for the applications discussed in Sections \ref{ch:offline} and \ref{hawc}
	illustrates how IceProd can be ported to other projects with minimal
	customization, which is facilitated by its Python code base.
	
	There are a couple of simple ways in which functionality can be extended:
	One is through the implementation of additional IceProd Modules as described in Section \ref{iceprodmodules}. 
	Another is by adding XML-RPC methods to the \emph{soapmon} module in order to provide a way for jobs to communicate with the server. 
	There are, of course, more intrusive ways of extending functionality, but those require a greater familiarity with the framework.

%
%
\section{Performance}
\begin{figure}
\begin{center}
\includegraphics[width=7.5cm,keepaspectratio]{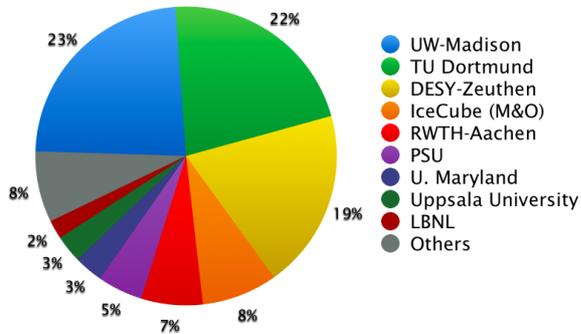}
\end{center}
\caption{Share of CPU resources contributed by members of the IceCube Collaboration towards simulation production. The relative contributions are integrated over the lifetime of the experiment. The size of the sector reflects both the size of the pool and how long a site has participated in simulation production.}
\label{fig:pie} 
\end{figure}

	Since its initial deployment in 2006, the IceProd framework has been instrumental in generating Monte Carlo simulations for the IceCube collaboration. 
	The IceCube Monte Carlo production has utilized more than three thousand CPU-core hours distributed between collaborating institutions at an increasing 
	rate and produced nearly two petabytes of data distributed between the two principal storage sites in the U.S.  
	and Germany.  Figure \ref{fig:pie} shows the relative share of CPU resources contributed towards simulation production. 
	The IceCube IceProd grid has grown from 8 sites to 25 over the years and incorporated new computing resources. Incorporating new sites is trivial since each set of daemons acts as a 
	volunteer that operates opportunistically on a set of job/tasks independent of other sites. There is no central manager that needs to scale with the number of computing sites. 
	The central database is the one component that does need to scale up and can also be a single point of failure. Plans to address this weakness will be discussed in Section \ref{future}.
	
	The IceProd framework has also been successfully used for the off-line processing of data collected from the IceCube detector over a 4-year period beginning in the Spring of 2010.
	This corresponds to 500 terabytes of data and over $3 \times 10^{11}$ event reconstructions.  
	Table \ref{tab:simprod-stats} summarizes the resources utilized by IceProd for simulation production and off-line processing.
\begin{table}  
\caption[IceCube resource utilization]{IceCube simulation production and off-line processing resource utilization. The production rate has steadily increased since initial deployment. The numbers reflect utilization of owned computing resources and opportunistic ones.}
\label{tab:simprod-stats}
	\centering 
	\begin{tabular}{l|r|r}\hline \hline
		&  Simulation & Off-line  \\
	\hline
	Computing centers	&  $25$ & 1 \\
	CPU-core time	&  $\sim 3000$ yr & $\sim 160$ yr \\ 
	CPU-cores           	&  $\sim 45000$ & 2000 \\
	No. of datasets		&  $2421$  & 5 \\
	No. of jobs		&  $1.6 \times 10^7$  & $1.5 \times 10^6$  \\
	No. of tasks		&  $2.3 \times 10^7$  & $1.5 \times 10^6$ \\
	Data volume    		&  $1.2$ PB & $0.5$ PB \\ \hline
	\end{tabular}
\end{table}

%
%
\section{Future Work}
\label{future}
      Development of IceProd is an ongoing effort. One important area of current development is the 
	  implementation of workflow management capabilities like HTCondor's
	  DAGMan but in a way that is independent of any batch system in order to optimize the use of 
	  specialized hardware and network topologies by running different job subtasks on different nodes. 
	  
	  Work is also ongoing on a second generation of IceProd designed to
	  be more robust and flexible. The database will be partially
	  distributed to prevent it from being a single point of failure and to better
	  handle higher loads. Caching of files will be more prevalent and
	  easier to implement to optimize bandwidth usage.  The JEP will be made
	  more versatile by executing ordinary scripts in addition to
	  modules. Tasks will become a fundamental part of the design rather than an added feature and will therefore be fully supported throughout the framework. 
	  Improvements in the new design are based on lessons learned from the first generation IceProd and provide a better foundation on which to continue development.

\section{Conclusions}
		IceProd has proven to be very successful for managing IceCube simulation
		production and data processing across a heterogeneous
		collection of individual grid sites and batch computing clusters.
				
		With few software dependencies, IceProd can be deployed and
		administered with little effort.
		It makes use of existing trusted grid technology and network protocols, which help
		to minimize security and data integrity concerns that are
		common to any software that depends heavily on communication through the Internet.

		Two important features in the design of this framework are the
		\emph{iceprod-modules}  and \emph{iceprod-server}  plugins, which allow
		users to easily extend the functionality of the code. The former
		provide an interface between the IceProd framework and user
		scripts and applications. The latter
		provide an interface that abstracts the details of job
		submission and management in different grid environments from the framework.
		IceProd contains a growing library of plugins that
		support most major grid and batch system protocols.

		Though it was originally developed for managing IceCube
		simulation production, IceProd is general enough for many types of grid applications and there are
		plans to make it generally available to the scientific
		community in the near future.

\input{ack}
The authors would like to also thank T. Weisgarber from the HAWC
collaboration for his contributions to IceProd development.

\appendix

\section*{Appendix}
The following is a comprehensive list of sites participating in IceCube Monte Carlo production:
Uppsala University (SweGrid), 
Stockholm University (SweGrid),
University of Alberta (WestGrid),
TU Dortmund (PHiDO, LIDO), 
Ruhr-Uni Bochum (LiDO),
University of Mainz,
Universit\'e Libre de Bruxelles/Vrije Universiteit Brussel,
Universiteit Gent (Trillian)
Southern University (LONI),
Pennsylvania State University (LIONX),
University of Wisconsin (CHTC, GLOW, NPX4),
Open Science Grid,  
RWTH Aachen University (EGI), 
Universit\"at Dortmund (EGI),
Deutsches Elektronen-Synchrotron (EGI, DESY),
Universit\"at Wuppertal (EGI), 
University of Delaware,
Lawrence Berkeley National Laboratory (PDSF, Dirac, Carver),
University of Maryland.

\bibliographystyle{elsarticle-num}
\bibliography{iceprod}

\end{document}

%% file: authors-elsv.tex
\author[Adelaide]{M.~G.~Aartsen}
\author[MadisonPAC]{R.~Abbasi}
\author[Zeuthen]{M.~Ackermann}
\author[Christchurch]{J.~Adams}
\author[Geneva]{J.~A.~Aguilar}
\author[MadisonPAC]{M.~Ahlers}
\author[Erlangen]{D.~Altmann}
\author[MadisonPAC]{C.~Arguelles}
\author[MadisonPAC]{J.~Auffenberg}
\author[Bartol]{X.~Bai\fnref{SouthDakota}}
% ...
\address[Adelaide]{School of Chemistry \& Physics, University of Adelaide, Adelaide SA, 5005 Australia}
% ...
\address[Christchurch]{Dept.~of Physics and Astronomy, University of Canterbury, Private Bag 4800, Christchurch, New Zealand}
% ...
\address[Erlangen]{Erlangen Centre for Astroparticle Physics, Friedrich-Alexander-Universit\"at Erlangen-N\"urnberg, D-91058 Erlangen, Germany}
\address[Geneva]{D\'epartement de physique nucl\'eaire et corpusculaire, Universit\'e de Gen\`eve, CH-1211 Gen\`eve, Switzerland}
% ...
\address[MadisonPAC]{Dept.~of Physics and Wisconsin IceCube Particle Astrophysics Center, University of Wisconsin, Madison, WI 53706, USA}
% ...
\address[Bartol]{Bartol Research Institute and Dept.~of Physics and Astronomy, University of Delaware, Newark, DE 19716, USA}
% ...
\address[Zeuthen]{DESY, D-15735 Zeuthen, Germany}
\fntext[SouthDakota]{Physics Department, South Dakota School of Mines and Technology, Rapid City, SD 57701, USA}
\fntext[Bari]{also Sezione INFN, Dipartimento di Fisica, I-70126, Bari, Italy}
\fntext[Goddard]{NASA Goddard Space Flight Center, Greenbelt, MD 20771, USA}

\cortext[cor1]{
	Corresponding author 
}
\cortext[cor2]{
	Principal corresponding author
}

%% file: ack.tex
\section*{Acknowledgements}
We acknowledge the support from the following agencies: 
U.S. National Science Foundation-Office of Polar Programs, 
U.S. National Science Foundation-Physics Division, 
University of Wisconsin Alumni Research Foundation, 
the Grid Laboratory Of Wisconsin (GLOW) grid infrastructure at the University of Wisconsin\textendash Madison, 
the Open Science Grid (OSG) grid infrastructure; 
U.S. Department of Energy, and National Energy Research Scientific Computing Center, 
the Louisiana Optical Network Initiative (LONI) grid computing resources; 
Natural Sciences and Engineering Research Council of Canada, WestGrid and Compute/Calcul Canada; 
Swedish Research Council, Swedish Polar Research Secretariat, 
Swedish National Infrastructure for Computing (SNIC), and Knut and Alice Wallenberg Foundation, Sweden; 
German Ministry for Education and Research (BMBF), 
Deutsche Forschungsgemeinschaft (DFG), 
Helmholtz Alliance for Astroparticle Physics (HAP), 
Research Department of Plasmas with Complex Interactions (Bochum), Germany; 
Fund for Scientific Research (FNRS-FWO), 
FWO Odysseus programme, 
Flanders Institute to encourage scientific and technological research in industry (IWT), 
Belgian Federal Science Policy Office (Belspo); 
University of Oxford, United Kingdom; 
Marsden Fund, New Zealand; 
Australian Research Council; 
Japan Society for Promotion of Science (JSPS); 
the Swiss National Science Foundation (SNSF), Switzerland; 
National Research Foundation of Korea (NRF); 
Danish National Research Foundation, Denmark (DNRF) 